\documentclass[twocolumn,showpacs,preprintnumbers,amsmath,amssymb,prb]{revtex4}

\usepackage{graphicx}
\usepackage{bm}

\begin{document}

\title{Andreev bound states in normal and ferromagnet/high-T$_c$ superconducting tunnel junctions}

\author{Mario Freamat and K.-W. Ng}
\affiliation{Department of Physics and Astronomy, University of Kentucky, Lexington,
KY 40506-0055, U.S.A.}

\date{\today}

\begin{abstract}
Ag/BSCCO and Fe/Ag/BSCCO planar tunnel junctions were constructed in order to
study experimentally the effect of an exchange potential on the spin polarized 
current transported through Andreev bound states appearing at the interface with 
a superconductor with broken time reversal pairing symmetry. The zero bias conductance 
peak (ZBCP) resulting from the Andreev bound states (ABS) is split into two symmetric 
peaks shifted at finite energies when the counterlectrode is normal. Four asymmetric 
peaks are observed when the ferromagnetic spin polarized charge reservoir is added, 
due to the combined effect of a spin-filtering exchange energy in the barrier, which 
is a spin dependent phenomenon, and the spin independent effect of a broken time reversal
symmetry (BTRS). The polarization in the iron layer leads to asymmetry. Due to the 
shift of ABS peaks to finite energies, the conductance at zero energy behaves as 
predicted by recent theoretical developments for pure $d$-wave junctions without 
Andreev reflections. 
\end{abstract}

\pacs{74.45.+c, 74.50.+r, 75.70.-i }\maketitle

\section{\label{sec:level1}Introduction\protect\\}
Tunneling conductance spectroscopy performed on d-wave
superconductor-insulator-counterelectrode junctions is one of the
most powerful experimental tools in the study of pairing properties of
high T$_c$ superconductors (SC) which are still incompletely understood.
The tunnel spectra provide a direct measurement of the energy resolved
density of states (DOS). One of the most suggestive spectral contributions
is due to the Cooper pairs transmitted into the SC as result of the low
energy Andreev reflections, leading to spectral features influenced by
the superconductive pairing symmetry\cite{CHu}. Since the Andreev
reflection is phase sensitive, the onset and amplitude of Andreev bound
states (ABS), manifested through the zero bias conductance peak (ZBCP),
are a signature of the symmetry of order parameter (OP). In particular,
for the case of cuprates, the d-wave symmetry of the order parameter was
experimentally probed\cite{Cov}$^-$\cite{Lee} and theoretically
modeled\cite{Kas}$^-$\cite{Far} by analyzing the behavior of ZBCP when the
current is injected under different angles of incidence in the ab-plane. The
phase of quasiparticles experiencing multiple Andreev reflections by lobes
of the OP with different sign may interfere constructively at low energies,
resulting in charge-current carrying ABS close to the surface, represented
by subgap DOS peaks in the tunneling spectra. The amplitude of the bound
state, and hence the height of the ZBCP, is maximum when the injection
is along the points of the OP, e.g. close from the [110] surface
when the symmetry is $d_{x^2+y^2}$.

However, while the time reversal symmetric $d$-wave OP predominates in the
bulk of the SC\cite{Bai}, a nodeless broken time reversal symmetry (BTRS)
state may appear\cite{Fog,Cov} at the junction interface. This is
consistent with a $d_{x^2+y^2}$+i$s$ combination. Experimentally, this
results in a split of the ZBCP proportional in magnitude and with an
onset temperature dependent on the strength of the $s$ subcomponent.
A similar influence has an $s$-wave conterlectrode, or may be induced
by proximity effect\cite{Sin}.

In recent years much experimental\cite{Hao,Saw,Lee} and theoretical\cite{Kas,Ste}
effort was directed toward understanding the effect of an exchange field on the
spin polarized transport in hybrid ferromagnet/$d$-wave SC structures.
The ZBCP splitting is in fact an energy  shift of the ABS peaks and it can
be correlated not only to BTRS, which is a spin independent phenomenon,
but also to processes magnetic in nature. In this case, the spin distribution
of the tunneling carriers becomes important. Thus, it was suggested\cite{Hao,Kas,Saw}
that a spin filtering junction barrier, like a ferromagnet insulator, may
also lead to splitting by determining an imbalance between the energy for up
and down spin components transported through the junction. As result, the
corresponding ABS peaks are shifted towards finite negative and positive
energies. This phenomenon is observable as a split ZBCP, with the splitting
magnitude dependent on the barrier strength and the spin filtering exchange
energy in the barrier. Moreover, if the counterelectrode is a ferromagnet,
the DOS spectrum is expected to be asymmetric due to the spin polarization
which consequently can be calculated from the ratio of the two unequal
peaks\cite{Hao}.

In this paper we report our measurements on a planar
Fe/Ag/Bi$_2$Sr$_2$CaCu$_2$O$_8$ (BSCCO) junction where the BTRS splitting
effect combines with a the spin dependent energy shift. Due to the Fe
polarization, the peaks are asymmetric with the holelike branch (E $<$ 0)
more prominent than the electronlike one. We compare the spectral
characteristics and their variation with temperature with an unpolarized
junction, Ag/BSCCO.

\section{Experiment and Discussion}
\begin{figure}
\includegraphics{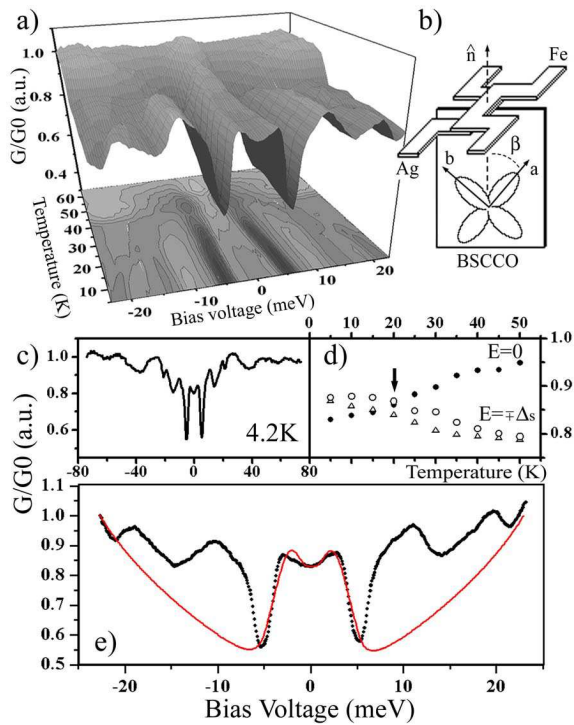}
\caption{\label{fig:F1} a) Variation with temperature of the $d$-wave subgap
	portion of the tunneling spectra obtained on Ag/BSCCO junction. The actual spectra
	are more V-shaped - common for pairing potentials with lines of nodes - and
	contain the edges of the $d$-wave gap (see inset c). Here we normalized by the
	spectrum at 100 K to flatten the ZBCP background, for more visibility.
	b) Junction layout. The Ag/BSCCO junction contains only the Ag layer. c) Complete
	spectrum at 4.2K, with $d$-wave gap coherence peaks visible. d) Temperature
	dependence of experimental conductance at E = 0, and at E = $\pm \Delta_s$
	($s$-wave gap). The splitting occurs at $\approx$ 20 K, where the curves intersect.
	e) Spectrum at 4.2 K with the BTK-type fit. The corresponding
	parameters are $\Delta_d$ = 33 meV, $\Delta_s$ = 2.4 meV, $U$ = 0, $h$ = 0
	(no exchange potential), $Z$ = 2.5, $\beta$ = 14$^\circ$, $\Gamma$ = 1.27 meV. For
	this $\beta$, a different behavior is expected for the conductance at
	E = 0. The discrepancy is explained in text.}
\end{figure}

Our junctions employ BSCCO monocrystals cleaved from samples prepared by the
self-flux method. The thin film metallic counterelectrodes were evaporated
normal on the ab-plane of the cuprate (Fig.~\ref{fig:F1}a). Two types of
junctions were built: Ag/BSCCO and Fe/Ag/BSCCO junctions. (The procedure
is presented in more detail elsewhere\cite{Mar}.) The thickness of Fe layer
was controlled to be $\approx$ 60\AA. For approximately this thickness and
junction orientation $\beta$=45$^\circ$, the amplitude of the ABS at zero bias
presents a maximum\cite{Far,Mar}. The 30\AA\ Ag layer in the ferromagnetic
junction was intended to decrease the junction interface strength which, if
too high, may suppress the spin splitting effect in the barrier. We
collected I-V data by standard 4-point measurements, from which the
differential conductance was calculated.
In our experiment the junction orientation in the ab-plane (angle $\beta$ in
Fig.~\ref{fig:F1}b) cannot be predetermined. Only some junctions show
the ZBCP, indicating a predominant injection with $\beta \neq$ 0.
In Fig.~\ref{fig:F1}a we present the temperature dependent
conductance spectra obtained on one of the Ag/BSCCO junctions. The
T$_c \approx$ 85 K and the ZBCP splits with the onset of the $s$-wave
subcomponent, at T$_s \approx$ 20 K, as shown in Fig.~\ref{fig:F1}d
where the splitting occurs at the intersection between conductance
curves at E = 0 and E = $\pm \Delta_s$. The splitting is symmetric and the
amplitude of the peaks is relatively large as in the unpolarized conditions
the spin-up and spin-down wave vectors of the quasiparticles are equal,
leading to a balanced, maximal contribution from Andreev reflections in both
branches of the spectrum.

In Fig.~\ref{fig:F1}e we fitted the subgap spectrum at T = 4.2 K, using a
BTK-type model \cite{Kas,Ste} modified for anisotropic pairing symmetries.
It calculates the DOS in terms of the probability amplitudes of different
events occurring at the junction interface. Considering the tunneling
conductance for the spin-up and spin-down quasiparticles separately, the
total conductance for the charge current incident under angle $\theta$, at
energy E, is $\sigma_q(E)=\sigma_{q\uparrow}+\sigma_{q\downarrow}$. The
fitting parameters are the $d$+i$s$-wave gaps $\Delta_{d,s}$, junction
orientation $\beta$, barrier strength $Z$ and exchange amplitude $U$,
and the ferromagnetic exchange potential $h$ of the counterlectrode. To
emulate the effect of temperature, a Gaussian smearing factor $\Gamma$
was introduced.

The fit results in a magnitude of 7.3\% for the $s$-wave subcomponent. The
peaks on the sides of the ZBCP are finite energy ABS\cite{Zar} and are
not discussed in this paper and are not included in the fit. The effect
of the BTRS on these peaks is also a shift at lower/higher energies
correlated with the ZBCP shift, looking like an increasing inner gap.
There is no spin dependent imbalance between the peaks resulting from the
ZBCP since there is no exchange field ($h$ = 0) and the two contributions
$\sigma_{q\uparrow}$, $\sigma_{q\downarrow}$ are equal. Recently Ref. 9
calculated the temperature dependence of the zero bias conductance
(ZBC, tunnel conductance at E = 0) in $d$ and $p$-wave SC structures.
According to this reference, the $d$-wave ZBC is expected to increase
with decreasing temperature at $\beta \neq$  0 and finite $Z$ due to
the enhanced ZBCP. As shown in Fig.~\ref{fig:F1}, in our case the ZBC
decreases monotonically with decreasing temperature, but this contradiction
is just apparent because of the splitting. Thus, our junction show a well
developed ZBCP at high temperatures, which is common for $\beta \neq$ 0
tunneling. If the junction were pure $d$-wave, the ZBCP would increase
with decreasing temperature, following the theory outlined in Ref. 9. We
explain the temperature dependence of ZBC by the effect of BTRS which
flattens and then splits the ZBCP at lower temperatures, leaving the ZBC
on the bottom of the inner gap to evolve as in the case of a $\beta$=0,
no ZBCP, junction. On the other hand, the conductance at E = $\pm \Delta_s$
behaves as Ref. 9 predicts for the ZBC of a junction with parameters
similar to ours: it varies proportional to the inverse temperature
(Fig.~\ref{fig:F1}d). Another possible explanation, i.e., the presence
of a spin-triplet pairing like p$_y$, is improbable since it
brakes the time reversal symmetry\cite{Ste} and hence the ZBCP would be
already split at T$_c$. This is not the case in our experiment where the
splitting occurs at T $\approx$ 20 K $<$ T${_c} \approx$ 85 K.
\begin{figure}
\includegraphics{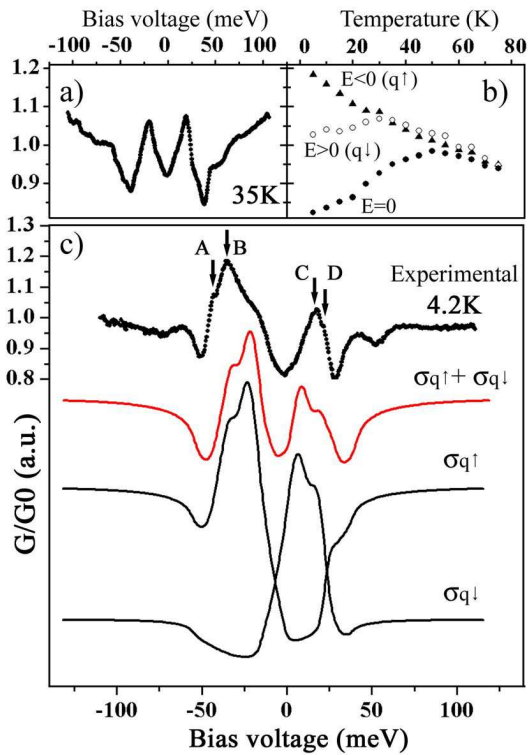}
\caption{\label{fig:F2} a) Conductance spectrum measured on Fe/Ag/BSCCO junction at
	higher temperature. The peaks are still symmetrical since the iron magnetization
	is in the plane of the thin film. b) Temperature dependence of
	conductance at E = 0 and at the positions of the ABS peaks shifted by the combined
	effect of BTRS state and spin splitting interaction in the barrier. e) Experimental
	spectrum at 4.2 K (dots). The A-B-C-D arrows indicate the 4 peak-like features
	attributed to the combined splitting. The BTK-type fit (continuous lines) includes
	the curves for spin up/down carriers (showing how the corresponding ABS move towards
	finite energies) and the total spectrum . The corresponding parameters are
	$\Delta_d$ = 45 meV, $\Delta_s$ = 7.2 meV, $U$ = 2.7 (or a difference of 5.4 between
	the barrier felt by the spin up and down currents), $h$ = 0.086, $\beta$ = 45$^\circ$,
	$\Gamma$ = 1.27 meV.}
\end{figure}
The presence of an exchange field in the Fe/Ag/BSCCO junction and the appearance
of a naturally induced\cite{Saw} exchange potential in the junction barrier,
combines with the BTRS to influence the ABS, leading to a four-peak asymmetric
splitting, marked in Fig.~\ref{fig:F2}c with ABCD arrows. The B-C splitting is due
to the spin filtering effect and the A-B, C-D splittings to the much smaller effect
of BTRS state. The interface exchange energy filters the spins by lowering the
barrier strength for the spin up current and raising it for spin up component,
so that the corresponding Andreev peaks (initially located at zero bias, in
the absence of BTRS) gain and loose energy, moving at different energy. This
phenomenon depends on the quasiparticle trajectories, with the trajectories normal
on the barrier experiencing the maximum spin splitting. As result, the ZBCP not
only splits, but the resulting peaks broadens with growing exchange potential $U$.
This broadening was observed in our experiment, as seen in Fig.~\ref{fig:F2}a,c.
The contribution of the barrier spin filtering leads to a peak-to-peak (B-C)
splitting of $\approx$ 27 meV, much larger than the $s$-wave gap,
$\Delta_s \approx$ 2.4 meV, which measures the splitting in case of Ag/BSCCO
junction. The bumps on the sides of the peaks are traces of the BTRS effect,
resulting in the four-fold splitting (A-B and C-D in Fig.~\ref{fig:F2}c). We fitted
the 4.2 K spectrum using the same model as above and obtained the parameters
indicated in the figure. The $s$-wave  subcomponent is small, $\approx$6\%,
and the ABS peaks are broad and high in amplitude, consistent with the
$\beta$=45$^\circ$ injection. The ZBCP splitting is asymmetric,
since B-C splitting is spin dependent and hence influenced by the ferromagnet
polarization. Consequently, the polarization can be estimated from the peak ratio,
which in our case gives P$\approx$35\%, measured at 4.2 K. For a clearer image
of this effect, we plotted separately the peaks corresponding to
$\sigma_{q\uparrow,(\downarrow)}$ shifted to lower (higher)
energies. We observed that the splitting appears immediate under T$_c$, but the
peaks are symmetric until down to $\approx$ 32 K (Fig.~\ref{fig:F2}a). This is
visible in Fig.~\ref{fig:F2}b where we plotted the temperature variation of
conductance at zero energy and at energies corresponding to the ingap maxima
(ABS peaks). One can see that the peak asymmetry appears only at lower
temperatures. This can be explained by the specific temperature dependence of
the ferromagnetism of iron very thin layer. It was observed\cite{Guo} that,
in function of layer thickness and temperature, the magnetization in thin
ferromagnetic films can experience a reorientation transition from in-plane
to perpendicular direction, due to the temperature dependence of perpendicular
anisotropy in ultrathin magnetic films. Consequently, while at higher temperatures
the peaks are symmetrical since the magnetization is mainly parallel with junction
interface, starting with $\approx$ 32 K the spins gradually switch to a
direction perpendicular on the barrier, into BSCCO ab-plane, so that the iron
polarization affects the ABS peaks symmetry.

It was proved\cite{Saw,Hir} that the ABS amplitude decreases with increasing
exchange field due to the sensitivity of Andreev reflections to spin polarization,
represented in the BTK-type models by a suppression in the Andreev
term coefficient. Ref. 9 showed that, at low temperature, this sensitivity is
independent of the strength of the barrier and the exchange potential can be
evaluated directly from the ZBC since the ABS appear for all quasiparticle
trajectories. As in the case of Ag/BSCCO junction, as seen in Fig.~\ref{fig:F2}b,
the ZBC variation with temperature conforms qualitatively to Ref. 9 curves
for $d$-wave junctions with $\beta$=0, in opposition to the expectation for
a junction like ours, exhibiting high amplitude Andreev peaks which rather suggest
a nodal orientation (the fit curve actually imposes $\beta$ = 45$^\circ$). As in
the case of Ag/BSCCO junction, this just apparently contradictory behavior can
be explained by the ABS shift to finite energies, so that the low energy DOS,
devoid of ABS contribution, is given mainly by single particle processes. As
seen in Fig.~\ref{fig:F2}b, the conductance of maximum ABS amplitude (spin up peak
shifted at lower energy E $<$ 0) behave as expected from the pure $d$-wave ZBC,
with the distortions due to polarization. At lower temperature, due to polarization,
the spin up distribution is enhanced in the detriment of the lowering spin down
peak, so that the conductance of ABS at E $>$ 0 first increases and then decreases
with decreasing temperature, opposite to the monotonically increasing E=$\Delta_s$
conductance in the Ag/BSCCO case. The special case of some spin-triplet pairing
is also excluded since our junctions were highly transparent (small $Z$), so
that the Andreev reflections are too strong to be suppressed by the exchange
interaction responsible at higher $Z$ for the lowering of ZBC with decreasing
temperature, e.g. when the pairing is $p_y$. Moreover, even if the split appears
immediately under T$_c$, as may be expected for the $p$-wave BTRS, the asymmetric
effect of polarization on the ABS peaks proves that the splitting cannot be
attributed to the BTRS state since this is spin independent.

\section{Conclusion}
In this paper we report the results of tunneling measurements on two types of
junctions, with and without an exchange potential. The ABS in the case of
Ag/BSCCO junctions is shifted at low temperature to finite energies, both
positively and negatively, due to a BTRS state, so splitting occurs and the
zero energy conductance decreases with lowering temperature as in the case
of junctions with orientation unfavorable to Andreev reflections. However,
when the conductance is measured at the position of the shifted ABS peaks, we
observed the dependency proper to the zero energy behavior of pure $d$-wave
junctions with orientation similar to our junction. In the case of Fe/Ag/BSCCO
junctions, the ABS are affected by the combined influence of the spin independent
BTRS state and the spin dependent exchange interaction in the barrier. Consequently,
the splitting is much larger than the s-wave gap, the spectrum show four peaks,
and the spectrum is asymmetric due to the polarization of the ferromagnetic
layer. Again, the conductance at zero energy decreases with decreasing temperature
since the ABS are displaced to finite energies. The behavior predicted for zero
energy conductance is identified at the new position of the spin up peak. However,
the spin down peak still decreases with temperature due to the DOS imbalance
induced by polarization in the iron layer. The polarization starts to affect
the peaks at a transition temperature corresponding to a switch of the iron
ultrathin film magnetization from in-plane to normal direction. While the ZBC
temperature dependency match the theoretical curves for a $p_y$-wave junction,
spectral features like the peak asymmetry exclude the possibility of a
spin-triplet presence.


\begin{thebibliography}{15}
\bibitem{CHu} C.R. Hu, Phys. Rev. Lett. {\bf 72}, 1526 (1994)
\bibitem{Bai} D.B Bailey, M. Sigrist, and R.B. Laughlin, Phys. Rev. B {\bf 55}, 15239 (1997)
\bibitem{Fog} M. Fogelström, D.Rainer, and J.A.Sauls, Phys. Rev. Lett. {\bf 79}, 281 (1997)
\bibitem{Cov} M. Covington, M. Aprili, E. Paraoanu, L. H. Greene, F. Xu, J. Zhu, and C. A. Mirkin, Phys. Rev. Lett. {\bf 79}, 277 (1997)
\bibitem{Sin} S. Sinha and K.-W. Ng, Phys. Rev. Lett. {\bf 80}, 1296 (1998)
\bibitem{Hao} X. Hao, J. S. Moodera, and R. Meservey, Phys. Rev. B {\bf 42}, 8235, (1990)
\bibitem{Saw} A. Sawa, S. Kashiwaya, H.Obara, H. Yamasaki, M. Koyanagi, and Y. Tanaka, Physica B {\bf 284-288}, 493 (2000)
\bibitem{Lee} K. Lee, S. Kim, B. Friedman, D. Cha, and I. Iguchid, Physica C {\bf 352}, 135 (2001)
\bibitem{Kas} S. Kashiwaya, Y. Tanaka, N. Yoshida, M.R. Beasley, Phys. Rev. B {\bf 60}, 3572 (1999)
\bibitem{Hir} T. Hirai, Y. Tanaka, N. Yoshida, Y. Asano, J. Inoue, and S. Kashiwaya, cond-mat/0210693
\bibitem{Ste} N. Stefanakis, Phys. Rev. B {\bf 64}, 224 502 (2001)
\bibitem{Zar} M. Zareyan, W. Belzig, and Yu. V. Nazarov, Phys. Rev. B {\bf 65}, 184 505 (2002)
\bibitem{Far} Z. Faraii and M. Zareyan, cond-mat/0304336
\bibitem{Mar} M. Freamat and K.-W. Ng, cond-mat/0301081
\bibitem{Guo} W. Guo, L. P. Shi, and D. L. Lin, Phys. Rev. B {\bf 62}, 14259, (2000) and references therein.
\end{thebibliography}
\end{document}